\documentclass[aps, reprint,showpacs,showkeys,pra]{revtex4-1}
\usepackage{dcolumn}
\usepackage[utf8]{inputenc}
\usepackage{amsmath}
\usepackage{amsfonts}
\usepackage{amssymb}
\usepackage{graphicx}
\usepackage{hyperref}
\usepackage{xcolor}

\newcommand{\mb}{\mathbf}
\newcommand{\rme}{\mathrm{e}}

\begin{document}

\author{Raul Corr\^ea}\email{raulcs@fisica.ufmg.br}\author{Pablo L. Saldanha}
\affiliation{Departamento de F\'isica, Universidade Federal de Minas Gerais, Caixa Postal 701, 30161-970, Belo Horizonte, MG, Brazil}

\title{Photon reflection by a quantum mirror: a wave function approach}

\begin{abstract}
We derive from first principles the momentum exchange between a photon and a quantum mirror upon reflection, by considering the boundary conditions imposed by the mirror surface on the photon wave equation. We show that the system generally ends up in an entangled state, unless the mirror position uncertainty is much smaller than the photon wavelength, when the mirror behaves classically. Our treatment leads us directly to the conclusion that the photon momentum has the known value $\hbar\mathbf{k}$. This implies that when the mirror is immersed in a dielectric medium the photon radiation pressure is proportional to the medium refractive index $n$. Our work thus contributes to the longstanding Abraham--Minkowski debate about the momentum of light in a medium. We interpret the result by associating the Minkowski momentum (which is proportional to $n$) with the canonical momentum of light, which appears naturally in quantum formulations. 
\end{abstract}



\maketitle

The advances in nanofabrication techniques have allowed extensive research on mechanical systems whose masses are small enough so that they can significantly exchange momentum with electromagnetic fields -- the so-called optomechanical systems. An important class of those systems are the optomechanical cavities, in which the mechanical system is described by a harmonic oscillator that couples with the light \cite{aspelmeyer14}. In some cases the cavity is composed by mirrors, and since the mid 1990s there were proposals of studying quantum properties of such mirrors, for instance quantum fluctuations \cite{jacobs94} and the construction of superposition states \cite{mancini97,bose99,marshall03,bassi05,steuernagel11}. Also, recently there have been many theoretical discussions on entanglement between vibration modes of mirrors in a quantum regime and light \cite{vitali07,paternostro07,hofer11}, while it has been experimentally reported for other mechanical systems \cite{palomaki13,cohen15}. 

The quantum mechanical treatment for a cavity quantum mirror interacting with photons can be based on a Hamiltonian formulation \cite{law95}, with a second quantization approach for light. But here we take a different path, by using a photon wave function approach to treat the reflection of a single photon by a quantum mirror. Our treatment is constructed from first principles and can be used to describe photons interacting with quantum mirrors in a cavity as well as a single photon reflection by a quantum mirror. The mirror imposes boundary conditions for the photon wave function at its surface and this naturally leads to the photon radiation pressure on the mirror, { which is associated} to the photon phase change upon reflection. We discuss the mirror-photon entanglement and its dependence on the relation between the photon momentum and the mirror momentum uncertainty. When the mirror is immersed in a medium with refractive index $n$, we show that the radiation pressure is proportional to $n$, which agrees with experiments performed with classical mirrors \cite{jones51,jones78}. We analyze this effect based on the association of the Minkowski momentum (which is proportional to $n$) with the canonical momentum of light \cite{barnett10a}, contributing to the long-standing Abraham-Minkowski debate about the momentum of light in material media \cite{pfeifer07,barnett10b}. 

\par We consider one photon in a paraxial beam state with arbitrary polarization that reaches a perfectly conducting plane surface (the mirror) and interacts with it during a finite time. The mirror surface is considered to be larger than the photon beam diameter and its description is quantized in the $z$ direction, which corresponds to the direction orthogonal to the surface plane. We consider the mirror initially in an arbitrary quantum state. We want to know the state of the system after the photon is completely reflected, while considering that the photon-mirror interaction occurs in a time scale much smaller than the one { by} the quantum mirror free evolution -- in such case the mirror wave function can be considered stationary during the reflection process. Our strategy to solve the problem is to make use of the linearity of both Schr\"odinger's and Maxwell's equations, to find the known solution of a monochromatic classical electromagnetic field being reflected by a fixed infinite plane mirror, and to construct the arbitrary solution from the superposition of those. In order to do this, we are going to describe both the mirror and the photon with wave functions. 

The mirror wave function is the traditional quantum mechanical wave function whose time evolution is described by the Schr\"odinger equation. For a classical perfectly fixed mirror, its quantum state can be approximated as a Dirac delta wave function in the position space, having a momentum uncertainty that tends to infinity. Therefore no matter how much momentum it exchanges with a photon, its wave function can only acquire a global phase. Since an arbitrary quantum state for the mirror can be decomposed in the position eigenfunctions and since the Schr\"odinger equation is linear, if we know what the interaction does to every position eigenfunction, we know what it does to an arbitrary mirror state.
 
We use the Bialynicki-Birula--Sipe wave function description for the photon \cite{birula94,sipe95,smith07,saldanha11}. This photon wave function is a complex vector function of the spatial and time coordinates that completely describes the quantum state of a photon. It can be decomposed in the eigenstates of the helicity operator $\hat{\sigma}$ in the following way:
\begin{align}
\mb{\Psi}(\mb{r},t)=\mb{\Psi}_+(\mb{r},t)+\mb{\Psi}_-(\mb{r},t),
\end{align}
where
\begin{align}
\mb{\Psi}_\pm(\mb{r},t)=\sqrt{\frac{\epsilon_0}{2}}\mb{E}_\pm(\mb{r},t)\pm i\sqrt{\frac{1}{2\mu_0}}\mb{B}_\pm(\mb{r},t).
\end{align}
$\varepsilon_0$ represents the electric permittivity and $\mu_0$ the magnetic permeability of free space. We have $\hat{\sigma}\mb{\Psi}_\pm=\pm\mb{\Psi}_\pm$ and the condition $\nabla\cdot\mb{\Psi}=0$ is imposed. The helicity eigenstates are associated to photons with circular polarizations and the photon electric and magnetic fields are given by $\mb{E}=\mathrm{Re}[\sqrt{2/\varepsilon_0}\mb{\Psi}(\mb{r},t)]=\mb{E}_++\mb{E}_-$ and $\mb{B}=\mathrm{Im}[\sqrt{2\mu_0}\hat{\sigma}\mb{\Psi}(\mb{r},t)]=\mb{B}_++\mb{B}_-$. By introducing the term $\mb{J}(\mb{r},t)$ accounting for the induced current in the medium due to the presence of the photon field, Maxwell's equations for the electromagnetic field in a medium can be recovered with the use of the photon wave equation \cite{saldanha11}
\begin{align}
 i\frac{\partial\mb{\Psi}(\mb{r},t)}{\partial t}=c\hat{\sigma}\nabla\times\mb{\Psi}(\mb{r},t)- i\frac{\mb{J}(\mb{r},t)}{\sqrt{2\epsilon_0}}.
\end{align}
This Maxwell wave equation determines the photon evolution, just like the Schr\"odinger equation does for a quantum massive particle. We note that to apply the second quantization procedure to the electromagnetic field in the presence of matter is an extremely difficult task \cite{glauber91,huttner92,scheel08,philbin10}. The boundary conditions imposed by the interface between different media and the interaction of the electromagnetic field with dispersive and absorptive media makes the quantization process to be very complicated. In this sense, the use of the Maxwell wave equation greatly simplifies the treatment in relation to the second quantization method when there is no absorption or emission of photons in the problem to be treated. This is the case in the present problem of the photon reflection by a quantum mirror. Since the photon equation is equivalent to the Maxwell's equations, a boundary conditions problem for a photon interacting with different media has the same solution as the one for a classical electromagnetic field. It is worth mentioning that the photon wave function formalism can be useful even when there is the absorption and generation of photons in a scattering process, as in the generation of entangled twin photons with parametric down conversion \cite{saldanha11,saldanha13}.

\begin{figure}
  \centering
    \includegraphics[width=0.5\textwidth]{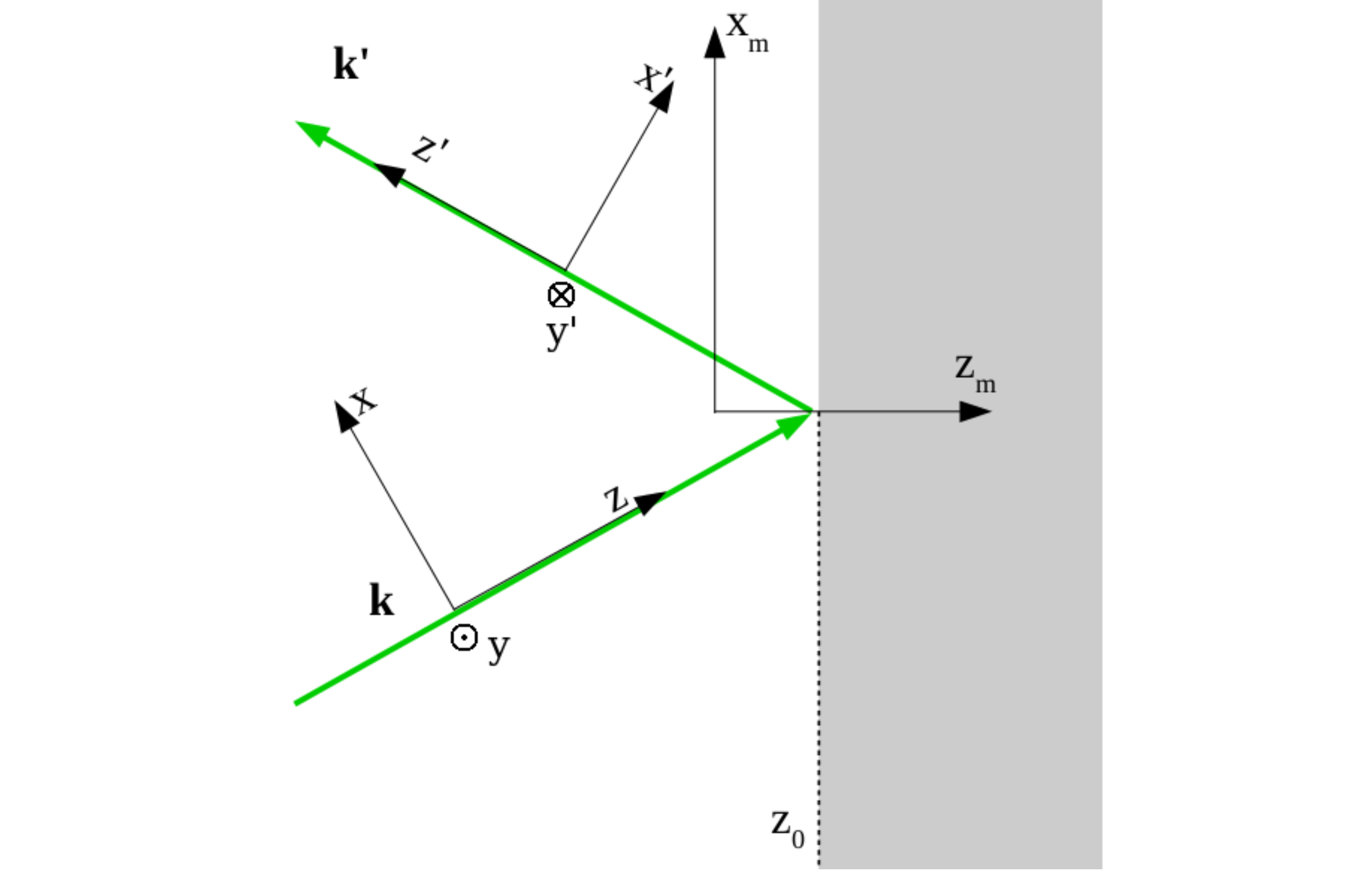}
  \caption{An electromagnetic field with wave vector $\mb{k}$ is reflected by a perfect mirror, resulting in a reflected wave with wave vector $\mb{k}'=\mb{k}-2(\mb{k}\cdot\mb{\hat{z}_m})\mb{\hat{z}_m}$.}\label{fig:mirror}
\end{figure}

\par Let $z_m=z_0$ be the plane of the mirror interface, with the region $z_m>z_0$ being a perfect conductor, as shown in Fig. \ref{fig:mirror}. It can be shown that the electromagnetic  field inside the conductor falls to zero rapidly and there is no field propagation inside the conductor \cite{jackson}. When the penetration depth is much smaller than the field wavelength, the field component parallel to the interface is approximately zero at its surface. Hence, by solving Maxwell's equations with this condition, we find that if there is a field ${\mb{E}_\pm(\mb{r},t)=\mb{\hat{u}}_{\mb{k}\pm} E_{0\mb{k}}\rme^{ i(\mb{k\cdot r}-\omega t)}}$ in the $z_m<z_0$ region, called the incident field, then there must be a field ${\mb{E'}_\mp(\mb{r},t)=-\mb{\hat{u}}_{\mb{k}'\mp} E_{0\mb{k}}\rme^{ i(\mb{k'\cdot r}-\omega t)}\rme^{2 i(\mb{k}\cdot\mb{\hat{z}_m})z_0}}$, called the reflected field, in the same region. This guarantees that, at the interface, the component of the electric field parallel to it is zero. Here ${\mb{k}'=\mb{k}-2(\mb{k}\cdot\mb{\hat{z}_m})\mb{\hat{z}_m}}$, where $\mb{\hat{z}_m}$ is the unit vector perpendicular to the surface of reflection and pointing inward the conductor, and $\mb{\hat{u}}_{\mb{k}\pm}$ are the circular polarization unit vectors. Given the reference frames $(x,y,z)$ and $(x',y',z')$ indicated in Fig. \ref{fig:mirror}, we define {$\mb{\hat{u}}_{\mb{k}\pm}=(\mb{\hat{x}}\pm i\mb{\hat{y}})/\sqrt{2}$ and $\mb{\hat{u}}_{\mb{k}'\pm}=(\mb{\hat{x}'}\pm i\mb{\hat{y}'})/\sqrt{2}$}. Hence, in the same spatial configuration, the boundary conditions demand that if the incident photon in this space is described by the wave function
\begin{align}
\mb{\Psi}_\pm(\mb{r},t)=\mb{\hat{u}}_{\mb{k}\pm} A \rme^{ i(\mb{k\cdot r}-\omega t)},\label{wfi}
\end{align}
in the region $z_m\leq z_0$, then in that same region there must be a reflected part of the the wave function given by
\begin{align}
\mb{\Psi'}_\mp(\mb{r},t)=-\mb{\hat{u}}_{\mb{k}'\mp} A \rme^{ i(\mb{k'\cdot r}-\omega t)}\rme^{2 i(\mb{k}\cdot\mb{\hat{z}_m})z_0},\label{wfr}
\end{align}
with ${\mb{k}'=\mb{k}-2(\mb{k}\cdot\mb{\hat{z}_m})\mb{\hat{z}_m}}$. From now on, we represent the state of a photon with wave vector $\mb{k}$ and helicity $\pm$ in the Dirac notation as $|\mb{k}\pm\rangle$. The reflection implies that for every $|\mb{k}\pm\rangle$ component of the field, there must be another component $|\mb{k}'\mp\rangle$ with the same amplitude and a phase difference $-\rme^{2 i(\mb{k}\cdot\mb{\hat{z}_m})z_0}$ in order to satisfy the boundary conditions. Of course, the wave function must be zero for $z_m>z_0$.

\par Up to now we have been dealing with plane waves, which extend themselves with the same amplitude through all times and all space, but in our problem the interaction takes place during a finite time and in a restricted region of the mirror. So in order to talk about before and after the reflection of the photon and to consider that the mirror surface is larger than the beam diameter, we make use of the superposition principle and allow the state of the photon to be a superposition of different wave vectors, therefore confining it in space and time. For an incident photon in a beam state
\begin{align}
|\psi\rangle=\int\psi(\mb{k})(c_{\mb{k}+}|\mb{k}+\rangle+c_{\mb{k}-}|\mb{k}-\rangle)\ d^3k,\label{awfi}
\end{align}
with $|c_{\mb{k}+}|^2+|c_{\mb{k}-}|^2=1$ for every $\mb{k}$ and ${\int|\psi(\mb{k})|^2 d^3k=1}$, our treatment implies that, apart from a global phase,  the reflected photon state must be
\begin{align}
|\psi'\rangle=\int\psi(\mb{k})(c_{\mb{k}-}|\mb{k'}+\rangle+c_{\mb{k}+}|\mb{k'}-\rangle)\rme^{2 i(\mb{k}\cdot\mb{\hat{z}_m})z_0}\ d^3k.\label{awfr}
\end{align}

\par We are finally ready to include the wave function for the $z$ position of the quantum mirror in the description. Eqs. (\ref{awfi}) and (\ref{awfr}) correspond to the situation of fixed mirror at the position $z_m=z_0$, that is, its state is described by the wave function $\langle z_m|z_0\rangle=\delta(z_m-z_0)$. Hence, for a mirror in an arbitrary state $|\phi\rangle$ with wave function $\phi(z_m)=\langle z_m|\phi\rangle$, the composite state of the system before the interaction is
\begin{align}
|\Psi\rangle=\int\psi(\mb{k})(c_{\mb{k}+}|\mb{k}+\rangle+c_{\mb{k}-}|\mb{k}-\rangle)\ d^3k\notag\\
\otimes\int \phi(z_m)|z_m\rangle\ dz_m,\label{mwfi}
\end{align}
which leads us to the state after the reflection
\begin{align}
|\Psi'\rangle=\iint &\psi(\mb{k})\phi(z_m)\rme^{2 i(\mb{k}\cdot\mb{\hat{z}_m})z_m}\notag\\
\times&(c_{\mb{k}-}|\mb{k'}+\rangle+c_{\mb{k}+}|\mb{k'}-\rangle)|z_m\rangle\ d^3k\ dz_m.\label{mwfr}
\end{align}
The state described in Eq. (\ref{mwfr}) has a different phase factor for each ket $|\mb{k}\pm\rangle|z_m\rangle$ of the composite system state. This phase depends on the eigenvalues $\mb{k}$ and $z_m$, which means that this is a non-separable, or entangled, state. So in general there are non-classical correlations between the photon and the quantum mirror after the photon reflection. 

\par We can also write the state of Eq. (\ref{mwfr}) in the linear momentum basis for the mirror $\{|p_m\rangle\}$, given that $\langle p_m|z_m\rangle=(2\pi\hbar)^{-\frac{1}{2}}\rme^{-ip_mz_m/\hbar}$. We are led to
\begin{align}
|\Psi'\rangle=\iint &\psi(\mb{k})\ \tilde{\phi}(p_m-2\hbar(\mb{k}\cdot\mb{\hat{z}_m}))\notag\\
\times&(c_{\mb{k}-}|\mb{k'}+\rangle+c_{\mb{k}+}|\mb{k'}-\rangle)|p_m\rangle\ d^3k\ dp_m,\label{mwfk}
\end{align}
where 
\begin{align}
\tilde{\phi}(p_m)=\langle p_m|\phi\rangle=(2\pi\hbar)^{-\frac{1}{2}}\int\phi(z_m)\rme^{-ip_mz_m/\hbar}dz_m.
\end{align}
From the above equations it is clear that every component $|\mb{k}\pm\rangle$ pushes the mirror by transferring a momentum $2\hbar(\mb{k}\cdot\mb{\hat{z}_m})$ to it. This is the exact necessary amount to conserve momentum, since the reflection simply inverts every photon wave vector component in the $z_m$ direction. It is interesting to note that we arrived at this result of the momentum transfer from the photon to the mirror simply by imposing boundary conditions on the photon reflection. No specification of the photon momentum was made. In other words, we can conclude that the photon momentum is given by the expression $\hbar\mb{k}$ simply by computing the momentum transfer to the mirror upon reflection and imposing momentum conservation.

\par We can analyze some classical limits of the quantum state of Eqs. (\ref{mwfr}) and (\ref{mwfk}). In the case when the mirror position wave function approximates a delta function, Eq. { (\ref{mwfr})} reduces to Eq. (\ref{awfr}) for the reflected photon, with the mirror state unaltered by the photon reflection. { In the view of Eq. (\ref{mwfk})}, this approximation is valid when the mirror momentum uncertainty is much larger than the momentum gained by the reflection of each $\mb{k}$ component of the photon state. Since $\Delta p_m\Delta z_m \sim \hbar$, this is equivalent to the mirror position uncertainty being much smaller than the wavelengths that compose the photon state. For an optical photon with average wavelength $\lambda\sim 500\text{nm}$, it means that the mirror should have a position uncertainty at least around $\Delta x\sim 10^{-7}{\text m}$ for significant entanglement effects to appear. It is important to note that the momentum transfer from the photon to the quantum mirror can be increased by a factor of $Q$ if the quantum mirror is one of the mirrors of a cavity with a quality factor $Q$. This is because the photon is reflected on average $Q$ times by the quantum mirror before leaving the cavity. Entanglement effects may arise in that way with, for instance, $\Delta x\sim 10^{-13}{\text m}$, with $Q\sim 10^6$. If the mirror is in the ground state of a quantum harmonic oscillator, the relation between its mass $m_0$, its resonance frequency $\omega_0$ and its position uncertainty is $\Delta x=\sqrt{\hbar/2m_0\omega_0}$ \cite{cohen}. In that sense, $m_0\omega_0$ gets smaller as $\Delta x$ grows larger, hence entanglement effects arise whenever $m_0\omega_0\sim 10^{-8}\text{kg/s}$ or smaller. A look at table II of \cite{aspelmeyer14} shows us that the suspended mirrors with smallest $m_0\omega_0$ have it of order $10^{-6} \text{kg/s}$ along with $Q\sim 10^6$ \cite{thompson08}, which is a bit far from the regime needed. So it is still not possible to effectively entangle spatial modes of a photon and a mirror upon reflection.

Another disentangled state limit occurs if the photon propagates as a nearly monochromatic beam along the direction $\mb{k_0}$ (but non-monochromatic enough so that the interaction is still much faster than any evolution due to the free mirror Hamiltonian). In a rough approximation, the final state of Eq. (\ref{mwfk}) is then almost disentangled and the mirror momentum wave function is displaced by $2\hbar(\mb{k}\cdot\mb{\hat{z}_m})$:
\begin{align}
|\Psi'\rangle\approx\int \psi(\mb{k})\ (c_{\mb{k_0}-}|\mb{k'}+\rangle+c_{\mb{k_0}+}|\mb{k'}-\rangle)\ d^3k\notag\\
\otimes\int\tilde{\phi}(p_m-2\hbar(\mb{k_0}\cdot\mb{\hat{z}_m}))|p_m\rangle\ dp_m.\label{beam}
\end{align}

Intermediate regimes account for mirror position uncertainty of the order of the average wavelength of non-monochromatic light, and those generally result in a non-separable state, as explicit in Eq. (\ref{mwfk}). Following the discussions of the above paragraphs, such regimes can in principle be achieved by engineering cavities with larger quality factors tuned to smaller light wavelengths. But since a high quality factor is associated with highly monochromatic light allowed in the cavity, present technology seems to be in a deadlock to try to probe this kind of entanglement. It is important to note, though, that if a cavity with the quantum mirror is in one arm of an interferometer, as proposed in \cite{marshall03}, entanglement between the photon and the mirror could be generated due to the superposition of the single photon propagating on each arm of the interferometer. The quantum superposition of the path in which the photon interacts with the mirror and transfer momentum to it with the path in which the photon does not interact and does not transfer momentum to the mirror may result in an entangled state. {  But an experimental realization of this proposal has not yet been accomplished.}

\par Now we address the historical Abraham--Minkowski debate, which concerns how the linear momentum carried by light behaves when it propagates through a dielectric medium \cite{pfeifer07,barnett10b}. The two apparently contrary views, due to Max Abraham and Hermann Minkowski, respectively indentify the momentum of the electromagnetic field either inversely or directly proportional to the refractive index of the medium. But it is important to note that when both the electromagnetic and material energy-momentum tensors are taken into account, the experimental predictions of Abraham's and Minkowski's formulations are equivalent \cite{penfield67, degroot72, pfeifer07}. Recently Barnett showed how the Abraham and Minkowski momenta can be associated to the kinetic and canonical momentum of the field respectively \cite{barnett10a,barnett10b}. It is clear that, according to equation (\ref{mwfk}), the momentum gained by every component $|p_m\rangle$ of the mirror is proportional to the wave vector component $\mb{k}$. If $\mb{\hat{u}}_{\mb{k}}$ is the unitary vector along the direction of $\mb{k}$, then we can write $\mb{k}=n(\omega/c)\mb{\hat{u}}_{\mb{k}}$, where $n$ is the refractive index of the medium in which the photon is propagating. Clearly, this corresponds to the Minkowski momentum for the photon, which is directly proportional to $n$. This behavior was observed in the experiments with classical light being reflected by classical mirrors immersed in dielectric media \cite{jones51,jones78}, and we present a fully quantum justification here. The answer to why is the Minkowski momentum that appears in this case lies on the fact that quantum mechanics is a Hamiltonian theory, based on canonical relations between position and momentum. The phase acquired upon reflection by the photon on Eq. (\ref{awfr}), which is dependent on the mirror position, is shared by both mirror and photon on Eq. (\ref{mwfr}). The canonical commutation relations in quantum mechanics define translation operators with the same form as these phase factors \cite{cohen}, hence turning those phase factors into momentum kicks, made explicit on each component of Eq. (\ref{mwfk}). It is natural then that our system will reveal the canonical momentum of the photon, which corresponds to the Minkowski momentum.

In summary, we have treated a single-photon reflection by a quantum mirror using the photon wave function formalism. This allowed us to treat the problem using boundary conditions on the photon wave equation instead of using the second quantization formalism for light. By computing the momentum transferred from the photon to the mirror, we concluded that a photon with wavevector $\mb{k}$ must have momentum $\hbar\mb{k}$ in order to achieve momentum conservation in the system, as expected. We also showed that in the case that the photon is not monochromatic and its average wavelength is of the order of the mirror position uncertainty, entanglement between them might appear with the reflection process. Finally we addressed a contribution to the Abraham-Minkowski debate by showing, with a quantum treatment from first principles, that the momentum transferred from a photon to a mirror immersed in a dielectric medium upon reflection is proportional to the medium refractive index. This result associates the photon momentum with the Minkowski momentum. This is natural given that the Minkowski momentum is associated with the canonical momentum of light, which is the momentum that should appear in a quantum treatment.

This work was supported by the Brazilian agencies CNPq and CAPES.


%

\end{document}